\documentstyle{article}

\def\be{\begin{eqnarray}}
\def\ee{\end{eqnarray}}
\def\nn{\nonumber}

\begin{document}

\hfill{hepth/9711194}\\
\phantom.\hfill{ITEP-TH-67/97}

\bigskip

\centerline{\Large{On Integrable Structure}}
\centerline{\Large{behind the Generalized WDVV Equations}}

\bigskip

\centerline{{\it A.Morozov}}

\centerline{{\it ITEP, Moscow}}

\bigskip

\bigskip

\centerline{ABSTRACT}

\bigskip

In the theory of quantum cohomologies the WDVV equations imply
integrability of the system $(I\partial_\mu - zC_\mu)\psi = 0$.
However, in generic situation
-- of which an example is provided by the Seiberg-Witten theory --
there is no distinguished direction (like $t^0$) in the moduli
space, and such equations for $\psi$ appear inconsistent.
Instead they are substituted by
$(C_\mu\partial_\nu - C_\nu\partial_\mu)\psi \sim
 (F_\mu\partial_\nu - F_\nu\partial_\mu)\psi = 0$,
where matrices $(F_\mu)_{\alpha\beta} =
\partial_\alpha \partial_\beta \partial_\mu F$.

\bigskip

\bigskip

\bigskip

\section{Quantum Cohomologies (a brief summary)}

The WDVV (Witten-\-Dijkgraaf-\-Verlinde-\-Verlinde) equations
\cite{WDVV} are an important ingredient of the theory of
quantum cohomologies ($2d$ topological $\sigma$-models)
and play a role in the formulation of mirror transform.
The central object in these studies is the {\it prepotential}:
a function of ``time''-variables\footnote{
If the prepotential is interpreted as the ``quantum'' deformation
of the generating function of intersection numbers on some manifold $M$
(a Gromov-Witten functional for $M$), the variables
$t^\alpha$ are associated with ``observables''$\phi_\alpha$
-- the elements of the cohomology ring $H^*(M)$.
Basically,
$$
F(t) =
\langle\ \exp
\left(\sum_{\alpha=1}^{dim\ H^*(M)} t^\alpha\phi_\alpha \right)
\rangle_0
$$
}
$F(t^\alpha)$, which satisfies the WDVV equations:
\be
C_\mu C_\nu = C_\nu C_\mu, \ \ \ \forall \mu,\nu.
\label{WDVV-eq}
\ee
Here $C_\mu$ are matrices,
\be
(C_\mu)^\alpha_\beta = \eta^{\alpha\gamma}(F_\mu)_{\gamma\beta},
\ \ \
(F_\mu)_{\alpha\beta} =  F_{\mu\alpha\beta} =
\frac{\partial^3 F}{\partial t^\alpha \partial t^\beta \partial t^\mu}
\label{3-der}
\ee
and the ``metric''
\be
\eta_{\alpha\beta}^{(0)} = (F_0)_{\alpha\beta} =
\frac{\partial^3 F}{\partial t^0\partial t^\alpha \partial t^\beta},
\ \ \
\eta^{\alpha\beta}\eta_{\beta\gamma} = \delta^\alpha_\gamma.
\label{metric-0}
\ee

In {\it conventional} theory of quantum cohomologies there is
a distinguished variable $t^0$ (associated with the unity
$\phi_0 = I$ in the ring $H^*(M)$), such that the metric
$\eta = \eta^{(0)} = F_0$ in (\ref{metric-0}) is constant:
\be
\partial\eta/\partial t^\alpha = 0
\label{metconst}
\ee
As a corollary, the
matrices $F_\mu$ and $C_\mu$ are independent of $t^0$.
In these sircumstances the set of WDVV equations (\ref{WDVV-eq})
together with the relations (\ref{3-der}) -- saying that the
``structure constants'' $C_\mu$ are essentially the third derivatives
of a single function $F(t)$ -- implies the consistency
condition
\be
\left[ {\cal D}_\mu(z), {\cal D}_\nu(z)\right] = 0
\ \ \
\forall \mu, \nu
\ee
for the set of differential equations \cite{DefHodge}
\be
{\cal D}_\mu(z) \psi_z =
\left(I\frac{\partial}{\partial t^\mu} - zC_\mu(t)\right)\psi_z(t) = 0
\ \ \ \
\left(\ \partial_\mu \psi_z^\alpha =
zC_{\mu\beta}^\alpha\psi_z^\beta\
\right)
\label{DefH}
\ee
with arbitrary ``spectral parameter'' $z$.

This reveals an integrable (Whitham-like) structure behind
the conventional WDVV equations. (Direct interpretation of
(\ref{DefH} is in terms of deformations of the Hodge structures
on Kahler manifolds.)

The ``Baker-Ahiezer vector-function'' $\psi_z(t)$ has
various intepretations.

First, \cite{Los,Giv},
as a function of $z$ it is a generating function of the
correlators,
linear in gravitational descendants (Morita-Mumford classes)
$c^n(\phi)$:
\be
\psi_{z,\rho}^\alpha(t) = \sum_{n=0}^\infty z^n
\ \langle c^n(\phi_\rho)\phi^\alpha e^{\sum_\beta t^\beta\phi_\beta}
\ \rangle_0
\ee
Thus $\psi_z$ is an important part of
the reconstruction of the {\it full spherical} prepotential
${\cal F}_0(t^\alpha_n) = \ \langle\
\exp \sum_{\alpha,n} t^\alpha_n c^n(\phi_\alpha) \ \rangle_0$,
$\#(\alpha) = dim\ H^*(M)$,
$n = 0,1,\ldots$ -- the generating function of correlators with
arbitrary number of descendants.
Original prepotential appears when all
descendant time-variables vanish:
$$
F(t_0^\alpha) = \left.{\cal F}_0(t^\alpha_n)\right|_{
t^\alpha_{n\geq 1} = 0}
$$
When descendants are included, $\psi_z$ satisfies a
hierarchy of quadratic equations
\be
\frac{\partial}{ \partial t^\mu_n}\psi_{z,\rho}^\alpha =
\eta^{\alpha\gamma} \langle c^m(\phi_\mu) \phi_\gamma
\phi_\beta\rangle \ \psi_{z,\rho}^\beta =
\psi_{z,\rho}^\beta \oint\frac{dy}{y^{m+1}}
\frac{\partial}{\partial t^\beta_0}\psi_{y,\mu}^\alpha
\ee
which is the ``quasiclassical'' limit of some {\it full}
(i.e. possessing a group-theory interpretation in the
spirit of \cite{gentau}) integrable hierarchy -- to which
the {\it full} prepotential, the generating function of all
correlators for all genera, is a solution.

Second, the function $\psi_z$ usually posseses
integral representations of the form
\be
\psi^\alpha_z(t) = \int_\Gamma \Omega^\alpha_z(t)
\ee
along some cicles on some manifold $\tilde M$ -- which is interpreted
as a {\it mirror} of $M$.
In concrete examples (see, for example, \cite{Giv})
this representation is implied by the hidden group-theory
structure behind integrable system (\ref{DefH}), which
allows to interpret $\psi_z$ as eigenfunctions of Casimir
operators. Such eigenfunctions are well known to possess
natural integrable representations, see \cite{Lapred}
and references therein.

Clarification of these constructions, associating some
(loop) algebra with a manifold, remains an interesting
open question.

\section{WDVV Equations in Seiberg-Witten Theory}

The WDVV-like equations are now known to arise in a somewhat
broader context than conventional quantum cohomologies.
Namely, one can relax the condition (\ref{metconst}) and study
the WDVV equations in the situation when there is no distinguished
modulus $t^0$ and no distinguished metric $\eta^{(0)}$.
Such situation arises, for example, in Seiberg-Witten theory
\cite{SW} of low-energy effective actions for $N=2$ SUSY Yang-Mills
models in four and five dimensions.
This theory is long known to involve integrable structures
\cite{int} and the prepotential (quasiclassical $\tau$-function)
theory \cite{prep}.
The WDVV-like equations arise in Seiberg-Witten theory in the
form \cite{MMM}:
\be
F_\mu F_\lambda^{-1} F_\nu = F_\nu F^{-1}_\lambda F_\mu,
\ \ \
\forall \lambda, \mu, \nu \nn \\
(F_\mu)_{\alpha\beta} =
\partial_\alpha\partial_\beta\partial_\mu F
\label{WDVV}
\ee
i.e. the role of the metric $\eta$ can be played by any
matrix $F_\lambda$ (actually, by any linear combination
of such matrices). Accordingly, the mutually commuting matrices
$C_\mu^{(\lambda)} = F^{-1}_\lambda F_\mu$,
\be
\left[ C_\mu^{(\lambda)},C_\nu^{(\lambda)}\right] = 0
\ \ \
\forall \mu, \nu
\ee
are now implicitely dependent on the choice of $\lambda$.\footnote{
To avoid confusion, the set (\ref{WDVV}) is not richer than
(\ref{WDVV-eq}), as it can seem: (\ref{WDVV}) with any {\it given}
$\lambda$ immediately implies the equations for all other
$\lambda$.
}

However, since generically there is no constant
(moduli-independent) matrix $F_\lambda$, the generalized
WDVV equations (\ref{WDVV}) no longer imply
(\ref{DefH}). This system of consistent equations is
instead substituted by\footnote{
As well as I understand such equations {\it per se} were studied
as an alternative to (\ref{DefH})
by B.Dubrovin (see ref.\cite{WDVV}) and other authors --
but in the context of conventional quantum cohomology theory,
with distinguished $t^0$-direction.}
\be
\left(\partial_\mu - C_\mu^{(\lambda)}\partial_\lambda\right)
\psi = 0
\ \ \
\forall \mu, \lambda
\label{int1}
\ee
or, in a more symmetric form,
\be
\left(F_\lambda \partial_\mu - F_\mu \partial_\lambda\right)
\psi = 0
\ \ \
\forall \mu, \lambda
\label{int2}
\ee

It is easy to see that the operators with different $\mu$ at the
l.h.s. commute with each other:
\be
\left[
\left(\partial_\mu - C_\mu^{(\lambda)}\partial_\lambda\right),
\left(\partial_\nu - C_\nu^{(\lambda)}\partial_\lambda\right)\right]
= \left[  C_\mu^{(\lambda)},C_\nu^{(\lambda)}\right]\partial_\lambda^2 +
\nn \\ +
 \left((\partial_\nu C_\mu^{(\lambda)}) - (\partial_\mu C_\nu^{(\lambda)})
 + C_\mu^{(\lambda)}  (\partial_\lambda C_\nu^{(\lambda)})  -
   C_\nu^{(\lambda)}  (\partial_\lambda C_\mu^{(\lambda)})
\right)\partial_\lambda
\ee
The first term at the r.h.s. vanishes due to the WDVV equations,
and the second one can be seen to vanish if the definition of
$C_\mu^{(\lambda)}$ is used together with the fact that $F_\mu$ are
matrices, consisting of third derivatives.
Eq.(\ref{int2}) is (\ref{int1}), multiplied by a matrix
$F_\lambda$ from the left.

It can still seem non-obvious that equations (\ref{int2})
are all consistent, i.e. that the vector $\psi$ can be chosen
in a $\lambda$-independent way. This follows from the
relation:
\be
F_\mu\partial_\nu - F_\nu\partial_\mu = \nn \\ =
F_\mu(\partial_\nu - C_\nu^{(\lambda)}\partial_\lambda) -
F_\nu(\partial_\mu - C_\mu^{(\lambda)}\partial_\lambda)
\ee

In order to return back from the generic system (\ref{int1})
to (\ref{DefH}), it is enough to choose
$\psi = e^{zt^0}\psi_z$, what is a self-consistent anzats
when all the $C_\mu^{(0)}$ are $t^0$-independent.

There is no spectral parameter in the system (\ref{int2}),
instead it is homogeneous (linear) in derivatives and
possesses many solutions. They can be formally represented
in the form:
\be
\psi(t) =
P\exp \int^t \left(dt^\mu C_\mu^{(\lambda)}\partial_\lambda\right)
= \left\{ I + \left(\int^t dt_1^\mu C_\mu^{(\lambda)}(t_1)\
+  \right.\right.    \nn \\ =  \left.\left. +
          \int^t dt_1^\mu C_\mu^{(\lambda)}(t_1)\partial_\lambda
\left(\int^{t_1} dt_2^\nu C_\nu^{(\lambda)}(t_2)\right) \ +  \ldots\right)
\partial_\lambda\right\} \tilde\psi(t)
\label{ser}
\ee
and one can choose, for example,
$\tilde\psi(t) = e^{zt^\lambda}$.
Then different terms of expansion of (\ref{ser}) in $z$ are
different solutions to (\ref{int1}).

As a simplest example, one can take
\be
F = \frac{1}{2}\left(t_1^2\log t_1 + t_2^2\log t_2 +
  (t_1-t_2)^2\log(t_1-t_2)\right)
\ee
which is the perturbative prepotential for
$SU(3)$ $N=2$ SYM model in $4d$. Then the first few solution to
(\ref{int2}) are:
\be
\psi = \left(\begin{array}{c} t_1 \\ t_2 \end{array}\right), \ \ \
\psi = \left(\begin{array}{c} t_1^2 - 2t_1t_2 \\ t_2^2 - 2t_1t_2
\end{array}\right), \nn \\
\psi = \left(\begin{array}{c} t_1^3 - 2t_1^2t_2 \\ -t_1^2t_2
\end{array}\right), \ \ \
\psi = \left(\begin{array}{c} t_1^3 - 2t_1^2t_2 \\ -t_1^2t_2
\end{array}\right), \ \ \
\psi = \left(\begin{array}{c} -t_1t_2^2 \\ t_2^3 - 2t_1t_2^2
\end{array}\right), \ \ \
\ee
By the way, $\psi^\alpha = t^\alpha$ is always a solution to
(\ref{int2}) -- this follows immediately from the definition of
$F_\mu$'s as the matrices of the 3-rd derivatives, which are
symmetric under permutations of indices.

\section{Conclusion and acknowledgements}

The purpose of this letter is to explain that appropriate
integrable structure on the moduli space exists behind the
generalized WDVV equations, i.e. existence of a constant
metric is not needed for such structure to emerge.
I do not touch here neither interpretation, nor implications
of this simple statement. They will be discussed elsewhere.

I appreciate illuminating discussions with A.Losev.
This work was partly supported by the grant
RFFI 96-15-96939.

\end{document}